\documentstyle[12pt]{article}

\def\be{\begin{equation}}
\def\ee{\end{equation}}
\def\bea{\begin{eqnarray}}
\def\eea{\end{eqnarray}}

\def\nab{\bigtriangledown}
\def\tpi{\tilde\Phi}

\begin{document}
\newcommand{\Psl}{\not\!\! P}
\newcommand{\dsl}{\not\! \partial}
\newcommand{\half}{{\textstyle\frac{1}{2}}}
\newcommand{\for}{{\textstyle\frac{1}{4}}}
\newcommand{\eqn}[1]{(\ref{#1})}
\def\la{\mathrel{\mathpalette\fun <}}
\def\a{\alpha}
\def\b{\beta}
\def\g{\gamma}\def\G{\Gamma}
\def\d{\delta}\def\D{\Delta}
\def\e{\epsilon}
\def\et{\eta}
\def\z{\zeta}
\def\t{\theta}\def\T{\Theta}
\def\l{\lambda}\def\L{\Lambda}
\def\m{\mu}
\def\f{\phi}\def\F{\Phi}
\def\n{\nu}
\def\p{\psi}\def\P{\Psi}
\def\r{\rho}
\def\s{\sigma}\def\S{\Sigma}
\def\ta{\tau}
\def\x{\chi}
\def\o{\omega}\def\O{\Omega}
\def\lagr{{\cal L}}
\def\cd{{\cal D}}
\def\k{\kappa}
\def\be{\begin{equation}}
\def\ee{\end{equation}}
\def\tz{\tilde z}
\def\tF{\tilde F}
\def\ri {\rightarrow}
\def\cf{{\cal F}}
\def\pa {\partial}
\def\bpa{\overline{\pa}}
\begin{flushright}
\today
\end{flushright}
\bigskip\bigskip
\begin{center}
{\large\bf T-DUALITY WITHOUT QUANTUM CORRECTIONS }
\vskip .9 cm

\vskip .9 cm
{\bf Enrique Alvarez
\footnote{  e-mail: enrial@daniel.ft.uam.es}\quad  
Javier Borlaf
\footnote{e-mail:javier@delta.ft.uam.es}\quad and
Jos\'e H. Le\'on
\footnote{e-mail:jhleon@delta.ft.uam.es}} 
 \vskip 0.05cm
Departamento de F\'{\i}sica Te\'orica, C-XI \\ 
Universidad Aut\'onoma de Madrid \\
28049 Madrid, Spain
\end{center}
\bigskip
\centerline{\bf ABSTRACT}
\bigskip
\begin{quote}
\ \ \ \ \ It is a well known fact that the classical (``Buscher'')
transformations of T-duality do receive, in general, quantum corrections.
It is interesting to check whether classical T-duality can be exact
as a quantum symmetry. The natural starting point is a $\sigma$-model
with N=4 world sheet supersymmetry. Remarkably, we find that (owing to the
fact that N=4 models with torsion
are {\it not} off-shell finite as quantum theories),the T-duality 
transformations for these models
get in general quantum corrections, with the only known exception of warped 
products of flat 
submanifolds or orbifolds thereof with other geometries.

\end{quote}

\newpage

\section{Introduction }

There has recently been some interest on quantum corrections to T-duality
transformations \cite{PALLA} (where by ``classical'' T-duality we will 
understand in all
 this work Buscher's transformations \cite{BU} (not including the
dilaton transformation).
In most works, however, the problem has been
set up as a pure quantum field theoretical one, forgetting conformal 
invariance, which is one of the major themes in T-duality as a string symmetry,
and, indeed, the origin of the dilaton transformation \cite{DIL}.
\par
Our point of view in this work has been, in a sense, the opposite one; that is,
start from a conformal invariant theory as well behaved quantum-mechanically 
as possible, and examine whether in this simplest of all interesting contexts,
T-duality can be implemented {\it without} quantum corrections, perhaps as
a Legendre transform \cite{ROC}.
\par
General sigma-models with N=4 (world-sheet) supersymmetry are widely believed
to be finite as quantum theories \cite{lag1}. 
( Ricci flat N=2, on the other hand, is not enough, 
because of the 
well-known
counterterms of Grisaru, van de Ven and Zanon \cite{GRI}).This set of models
is  then the most promising starting point for our purposes.

\par
Although it is expected on general grounds that the dual model is physically
equivalent to the original one, it is by now quite clear that N=4
supersymmetry will not always be manifest in the dual model \cite{GRIEGOS}.
One can classify all isometries in two types, translational and rotational, 
whose essential distintion is whether or not they have fixed points.
One can be slightly more precise in four dimensional target spaces, where
a translational isometry is by definition such that the (antisymmetric part
of) the covariant derivative of the Killing vector is (anti) self-dual:
\be
\nabla_{\mu}k_{\nu} = {\pm}{1\over 2}\sqrt{|g|} \epsilon_{\mu\nu}\,^{\rho\sigma}\nabla_{\rho}k_{\sigma}
\ee
When the isometry possesses fixed points (which appear as zeros of the 
Killing vector field), supersymmetry can still be preserved if the rank
of the matrix $M_{\mu\nu}\equiv \nabla_{\mu}k_{\nu}$ at the fixed point
is equal to 4 (nut), and the isometry is translational. On the other
hand, when the corresponding rank is 2 (bolt), manifest supersymmetry is lost.

\par
But even if the problem is restricted to translational isometries (without
fixed points), where N=4 supersymmetry is expected to be manifest in the
dual model,a potential paradox inmediatly comes to mind. The general framework
of N=4 does not force $g_{00} = 1$ in adapted coordinates to the Killing
$k = \partial_0$. Buscher's formulas then imply a dilaton in
the dual model $\phi = -{1\over 2} log \, g_{00}$, spoiling at least the
off-shell finiteness of the dual theory.
\par

If the gauging procedure \cite{GPR} is used, one can
regulate the determinant
coming from the integration of the gauge fields in such a way
that the correct dilaton is obtained \cite{BU}.
\par
The auxiliar gauged theory corresponding to a general (1,1) model, reads:

\be
\label{gauged}
S_{auxiliar}\, =\, -{1\over2\pi}\int d^2 z d^2 \theta ((g_{\mu\nu} +
B_{\mu\nu}) 
\nab_{+}X^{\mu} \nab_{-}X^{\nu} - \tilde X (D_{+}V_{-} + D_{-}V_{+}))
\ee
where the notation of reference \cite{mrocek}has been used, together with  
$\nab_{\pm}X^{\mu} \equiv  D_{\pm}X^{\mu} + k^{\mu}A_{\pm}$,
$A_{\pm}$ being
the gauge superfields ,$k^{\mu}$ the Killing vector, and $\tilde X$ the
Lagrange supermultiplier.

Performing now the (gaussian) integral over the gauge superfields the 
standard dual path integral is obtained up to the term :
\be
\label{dilaton}
\int DA_{+}DA_{-}exp(-{1\over2}\int d^2 z d^2 \theta g_{00}(X^i)
A_{+}A_{-} )
\ee
Writing the above path integral in components and integrating the part
corresponding to 
the gauge supermultiplet  we get two determinants (one of them coming from
the
bosonic components  and the other  from the fermionic ones)
 which naively cancel.

We know, however, that we can not trust this formal argument, because of
the fact that the beta funcions of both the metric and the antisymmetric 
tensor remain the same
as in the purely bosonic model, for which a dilaton transformation
is neccessary.

\par
In order to possess N=4 supersymmetry, in models without torsion
the target space has to be endowed with a hyperk\"{a}hler structure.
\cite{lag1}. It is rather clear that if we want the dual model to have
also manifest N=4 supersymmetry, the Lie derivative
of the complex structures with respect to the Killing must be zero.

\be
\label{triho} 
{\cal L}_{k}{J^{(X)}}\,=\,0
\ee

This
implies that in adapted
 coordinates the complex structures are independent of the cyclic coordinates.
When this condition is satisfied, the isometry is said to be triholomorphic.
In 4-dimensional target spaces triholomorphicity is equivalent
to (anti)self-duality \cite{Gibb}.
\par
Although we are interested in target spaces of arbitrary dimensions, we shall 
make our considerations explicit for dimension 4, because in this case
the general form of the metric tensor for hyperk\"ahler manifolds with 
one translational Killing
isometry is \cite{Gibb}

\be
\label{hyper}
ds^2 \, = \, V(d\tau + w_i dx^i)^2 + V^{-1}(dx^i dx^i)
\ee

with the conditions 

\be
\partial_{[j} w_{k]} \,= \, {(\pm 1)\over2}\epsilon_{jki}
\partial_i V^{-1}
\ee
in adapted coordinates to the Killing vector ${\pa\over\pa \tau}$.
\par
General N=4 models are widely believed to be finite to all orders in
perturbation theory \cite{Howe}, but only when the torsion vanishes
they are off-shell finite as well( \cite{lag1}\cite{ht} ). In order to stay 
in as firm a ground as possible, we shall then restrict our attention to this subset
of finite models.

\section{Translational duality in general four-dimensional N=4 models}

Buscher's formulas for the dual background yield:
\bea
\label{dual}
& d\tilde s ^2 \, = \, V^{-1}(d\tilde\tau ^2 + d\tilde x^i d\tilde x^i ) \\
& \tilde b \, = \, 2 w_i d\tilde\tau \wedge d\tilde x^i 
\eea

It has all N=4 supersymmetries
manifest because we have dualized with respect to a triholomorphic
isometry \cite{Hass, Sfet}.
Standard wisdom would then suggest
\cite{Howe} it to be finite as well. To be specific,
 the dual left-right  K\"{a}hler forms are 
\bea
\label{complex}
&& {\tilde J}^{i }_{-}\,=\, V^{-1}(\pm d\tilde\tau \wedge d\tilde x^i - 
{1\over2}\epsilon^{ijk} d\tilde x^j \wedge d\tilde x^k ) \\
&& {\tilde J}^{i}_{+}\, = \, V^{-1}(\mp d\tilde\tau \wedge d\tilde x^i -
{1\over2}\epsilon^{ijk} d\tilde x^j \wedge d\tilde x^k) 
\eea

A straightforward computation of the beta function corresponding to the
metric tensor \cite{Ketov,Zachos} gives, however, a non-zero result, namely:

\be
\label{Ricci}
\beta_{\mu\nu}\, = \, \tilde R_{\mu\nu}^+ \,= \,
- 2\tilde\nab_{\mu}\tilde\nab_{\nu}\tpi \, \neq \, 0
\ee

where $\tilde R^+\,_{\mu\nu\sigma\rho}\,$ is the Riemann curvature for the
connection 
\be
\Gamma_{\rho\sigma}^{\mu}\equiv \{\,_{\rho\sigma}^{\mu} \} + 
T_{\rho\sigma}^{\mu};
\ee
the torsion is defined through 
$\tilde T_{\mu\nu\sigma} \equiv {1\over2} (
\partial_{\mu}\tilde b_{\nu\sigma} + \partial_{\sigma}\tilde b_{\mu\nu} +
\partial_{\nu}\tilde b_{\sigma\mu} )$, and the dilaton is given in 
terms of the metric function by $\tpi \equiv {1\over 2} log\,V$.
\par
According to stardard wisdom (\cite{ht}), this means that the dual model
is indeed finite, but only on-shell; that is, after a field redefinition.

\par
The fact that a dilaton is needed in the dual model in order for it to be one-
loop conformally invariant means, of course, that the (manifestly) N=4 world-sheet
supersymmetric model given by \eqn{dual} is not conformally invariant by itself.
\par
It is not known , however,whether the preceding one-loop dilaton 
correction is enough
to ensure conformal invariance to all orders.

\section{Duality under translational isometries in N=4 models 
with metric of the warped product type }

In this section, the sigma model will be assumed to have n commuting 
isometries, $k_a$.
In adapted coordinates,
$k_a \equiv {\partial\over\partial y^{a}}$, the most general sigma model reads:
\bea
S = {1\over 2\pi} \int d^2 z d^2\theta E_{ab}(X) D_+  Y^a D_- Y^b
+ g_{ij}(X) D_+ X^i D_- X^j +  \nonumber\\
 F(X)_{ai}D_+ Y^a D_- X^i
+ F(X)_{ia}D_+ X^i D_- Y^a 
\eea
(denoting by $x^i$, i= n+1,...,d, the coordinates not adapted to the 
isometries.)
\par
It has been shown in the preceding section that even in the restricted N=4 
framework, models with torsion are not always finite. The original model will
be torsion free when $B_{ab}\equiv E_{[ab]} = F_{[ai]} = 0$ , and the dual model 
when the ``mixed'' terms
$F_{ai} = F_{ia} =0$; we further define
the matrix $G \equiv G_{ab}\equiv E_{(ab)}$.   
\par
The Ricci tensor for the above models reads, under those conditions:
\be
R_{ab} = {1\over 2} g^{ij} (G \Box\,_{ij} G + {1\over 2}
\partial_k G G^{-1}\partial^k G)\,_{ab}
\ee
\be
R_{ai} = 0
\ee
\be
R_{ij}=\widehat{R}\,_{ij} + {1\over 2} tr (\Box \,_{ij} G
+ {1\over 2} G^{-1} \partial_i G G^{-1} \partial_j G)
\ee

where a caret over an object means that it has been computed with
the quotient metric, $G_{ij}(x)$, and the operator $\Box\,_{ij}$
is defined by
\be
( G \Box\,_{ij} G)\,_{ab} \equiv \hat{\nabla}\,_i \hat{\nabla}\,_j
G_{ab} - {1\over 2}( \partial_i G G^{-1} \partial_j G + \partial_j G 
G^{-1} \partial_i G)\,_{ab}
\ee
\par
In the simple setting considered in this section, in which there is no
torsion neither in the initial, nor in the dual model, the dual metric
is obtained by the transformation $G \rightarrow G^{-1}$ (this being actually
the reason for introducing the convenient ``covariant'' operator
$\Box\,_{ij}$, transforming as $ \Box_{ij} G^{-1} = - G (\Box_{ij}G) G^{-1}$).
\par

Ricci flatness of both the original and the dual model (a neccessary condition
for them to be hyperk\"ahler) leads to
\be
tr \, G^{-1} \partial_i G = 0
\ee
which means that the corresponding dilaton $det \, G = const.$
\par
Conformal invariance then reduces to the conditions
\be
(g^{ij} \Box\,_{ij} G)_{ab} =0
\ee
\be
\widehat{R}\,_{ij} = - {1\over 4} tr G^{-1} \partial_i G G^{-1} \partial_j G
\ee
\par
As a particular instance of the above, four-dimensional hyperk\"ahler
target spaces admit a further triholomorphic Killing vector provided 
the metric can be written in coordinates adapted to the Killings,
${\partial\over\partial\tau}$ and ${\partial\over\partial z}$
as:
\bea
ds^2 = V(x,y)(d\tau)^2 + 2\omega(x,y) V(x,y)d\tau dz + \nonumber \\
(V(x,y)\omega(x,y)^2 + V^{-1}(x,y))dz^2 + V^{-1}(x,y) (dx^2 + dy^2)
\eea
and
\be
\partial_x V^{-1} = \pm \partial_y \omega
\ee
\be
\partial_y V^{-1} = \mp \partial_x \omega
\ee
It is curious to notice that the above equations can be interpreted as the
Cauchy-Riemann conditions for the analytic function $ f (x\pm i y)\equiv 
V^{-1} 
\pm i \omega$.
On the other hand, it is not difficult to show that only flat manifolds admit
further holomorphic isometries.
\par
In our previous notation, $det \, G = 1$, and moreover $G^{-1} = G$ up 
to relabellings of the coordinates, conveying the fact
 that the model is geometrically
T-self-dual (up to boundary conditions; this includes the usual $R\rightarrow
{1\over R}$ transformations in the toroidal case).

\section{Conclusions}
We have obtained the rather surprising result that T-duality does not
act ``classically'',(i.e., without quantum corrections), even in
the simplest of all sigma-models, namely those which enjoy N=4 world-sheet
supersymmetry.
\par
The reason for that stems from the fact that (as we have
explicitly checked), N=4 models with torsion
are not always conformally invariant to all loops.

\par
The only physical
situation we have found in which T-duality acts as a {\it classical} 
symmetry is the one in which the metric correspond
to a warped product. This includes as particular examples, (warped) 
products of tori (orbifolds) with other manifolds.
\par
Actually, even in four dimensional target spacetimes, we have not been
able to prove that those are the {\it only} allowed classical situations.
The further study of the general class of N=4 models with torsion
in arbitrary target space dimension is obviously one of the outstanding open
problems in this context.

\par

\section{ Acknowledgements.}
We have benefited very much from correspondence with Gerardo Aldaz\'abal
and Faheem Hussain. We are also grateful for clarifying remarks by G. Papadopoulos
and U. Lindstr\"om on a first version of this paper.
Our work has been partially supported by AEN/96/1664 and by
a CAM grant (JB) and a MUTIS fellowship from AECI(JHL).


\begin{thebibliography}{100}
\bibitem{BU} T.H. Buscher, Phys. Lett. B194 (1987) 51; B201 (1988) 466.
\bibitem{TSEY} A.A.Tseytlin, Mod.Phys.lett. A6 (1991) 1721.
\bibitem{PALLA} J.Balog, P.Forgacs, Z.Horvath, L.Palla:
Nucl.Phys.B(Proc.Suppl.) 49(1996) 16;\\ Phys. lettB. 388:121,(1996);
hep-th/9609198.

\bibitem{DIL} E. \'Alvarez and M.A.R. Osorio, Phys. Rev, D40
(1989) 1150. \\ 
P. Ginsparg and C. Vafa, Nucl. Phys. B269 (1987) 414. \\
E. Alvarez and Y. Kubyshin,hep-th/9610032
\bibitem{ROC} N.J.Hitchin, A.Karlhede, U.Lindstr\"{o}n and M. R\v{o}cek: 
Comm. Math. Phys 108,535 (1987);\\ M.R\v{o}cek and E. Verlinde, 
Nucl.Phys.B373,630(1992).


\bibitem{ht}C. Hull and P. Townsend, Nucl. Phys.B274 (1986),349
\bibitem{GRI} M.T.Grisaru, A.E.M.Van de ven and D.Zanon: Nucl.Phys. 
B277(1986)388;\\
Nucl.Phys. B277(1986) 409; Nucl.Phys. B287(1987) 189.
\bibitem{JACK} I.Jack, D.R.T.Jones and J.Panvel: hep-th/9311117.
\bibitem{mrocek}
Introduction to Supersymmetry. Martin R\v{o}cek, Stony Brook NY 11794-3840. 
\bibitem{GRIEGOS} E. Bergshoeff, R. Kallosh and T. Ortin, Phys. Rev D51
(1995) 3003\\
I. Bakas, Phys. Lett B343 (1995)103\\
I. Bakas and K. Sfetsos, Phys. Lett.B349(1995),448\\
S.F. Hassan, Nucl. Phys. B454 (1995)86;B460(1996),362\\
E. Alvarez, L Alvarez-Gaume, I. Bakas, Nucl. Phys. B457 (1995),3;Nucl.
Phys. (Proc. Suppl).46 (1996),16\\
K. Sfetsos,Nucl. Phys. B463(1996)33\\
E. Kiritsis, C. Kounnas and D. L\"{u}st: Int.J.Mod.Phys.A9.1361 (1994). 




\bibitem{GSW} M.B. Green, J.H. Schwarz and E. Witten. - "Superstring 
Theory." (Cambridge Univ. Press, Cambridge, 1987).

\bibitem{AAGL2} E. \'Alvarez, L. \'Alvarez-Gaum\'e and Y. Lozano, 
Nucl. Phys. B (Proc. Suppl.) 41 (1995) 1. 
\bibitem{GPR} A. Giveon, M. Porrati and E. Rabinovici, Phys. Rep. 244 
(1994) 77.\\ 
E. Alvarez, L. Alvarez-Gaum\'e, J.L.F. Barb\'on and Y. Lozano, Nucl. Phys.
B.
415(1994),71.
\bibitem{lag1}
L.Alvarez-Gaum\'e and D.Z.Freedman : Comm.Math.Phys. 80(1981)443.
\bibitem{Gibb}
G.W.Gibbons and P.J.Ruback : Comm.Math.Phys.115(1988)267
\bibitem{Hass}
S.F.Hassan : Nucl. Phys. B460.362(1996).
\bibitem{Sfet}
I.Bakas and K.Sfetsos : Phys.lettB 349,448,(1995).
\bibitem{Howe}
P.S.Howe and G.Papadopoulos : Nucl.Phys.B289(1987)264
\bibitem{hull1}
C. Hull: Nucl. Phys.B260(1985) 182. 
\bibitem{Ketov}
S.V.Ketov : Nucl.Phys.B294(1987)813
\bibitem{Zachos}
E.Braaten, Th.L.Curtright and C.K.Zachos : Nucl.Phys.B260(1985)630
\bibitem{Ger}
G.Aldazabal, F.Hussain and R.Zhang : ICTP preprint IC/86/400
\bibitem{harmonic} 
E. Ivanov: Phys. LettB356,239,1995.\\
A.Galperin, E.Ivanov, V.Ogievetskii and E.Sokatchev:
Class.Quant.Grav 2,617,1985;\\
Class.Quant.Grav 2,601,1985.

\end{thebibliography}
\end{document}